\documentstyle[epsfig]{mn}{}
\begin{document}

\def\mpc{h^{-1} {\rm{Mpc}}}
\def\up{h^{-3} {\rm{Mpc^3}}}
\def\uk{h {\rm{Mpc^{-1}}}}
\def\lsim{\mathrel{\hbox{\rlap{\hbox{\lower4pt\hbox{$\sim$}}}\hbox{$<$}}}}
\def\gsim{\mathrel{\hbox{\rlap{\hbox{\lower4pt\hbox{$\sim$}}}\hbox{$>$}}}}

\title{Galaxy groups in the 2dF redshift survey: The catalogue}

\author[M. Merch\'an \& A. Zandivarez]
{Manuel Merch\'an \& Ariel Zandivarez \\
Grupo de Investigaciones en Astronom\'{\i}a Te\'orica y Experimental, 
IATE, Observatorio Astron\'omico, Laprida 854, C\'ordoba, Argentina \\
Consejo de Investigaciones Cient\'{\i}ficas y T\'ecnicas de la Rep\'ublica 
Argentina }
\date{\today}

\maketitle 

\begin{abstract}
We construct a galaxy groups catalogue from the public 100K data release of 
the 2dF
galaxy redshift survey. The group identification is carried out 
using a slightly modified version of the group finding algorithm developed 
by Huchra \& Geller.
Several tests using mock catalogues allow us to find the optimal conditions
to increase the reliability of the final group sample. A minimum number of
4 members, an outer number density enhancement of 80 and a linking
radial cutoff of $200 \ km \ sec^{-1}$, are the best obtained values from 
the analysis. Using these parameters, approximately 90\% of groups
identified in real space have a redshift space counterpart. On the other hand 
the level of contamination in redshift space reaches to 30 \% including a 
$\sim 6\%$ of artificial groups and $\sim 24\%$ of groups associated with 
binaries or triplets in real space.
The final sample comprise 2209 galaxy groups covering the sky region 
described by Colless et al. spanning over the redshift range of $0.003 \leq z 
\leq 0.25$ with a mean redshift of 0.1. 
\end{abstract}


\section{Introduction} 
The study of galaxy groups is a very interesting area
of research because these density fluctuations lay between galaxies and
clusters of galaxies and may provide important clues to galaxy formation. 

Since the appearing of large galaxy redshift surveys more reliable 
detection of group of galaxies have been possible. Two pioneer works
develop the main identification algorithms. The first was introduced by Huchra
\& Geller (1992) and the second was proposed by Nolthenius \& White (1987);
both are friends-of-friends algorithms, differing only in the scaling
of the linking lengths. Frederic (1995) perform an extensive study of these
algorithms using N-body simulations, concluding that neither of them is 
intrinsically superior to the other, and the choice of one of them depends
on the purpose for which groups are to be studied. 
Recently several catalogues have
been constructed from different large galaxy redshift surveys using these
algorithms. Merch\'an et al. (2000) use the Updated Zwicky Catalogue 
(Falco et al. 1999)
generating a sample of 517 galaxy groups. Using the Nearby Optical Galaxy
Sample, Giuricin et al. (2000) construct one of the large  catalogue of 
loose groups and Tucker et al. (2000) found 1495 groups in the Las
Campanas Redshift Survey. Finally, Ramella et al. (2002) generate a group catalogue of 1168 members from the combination of the 
Updated Zwicky Catalogue and the Southern Sky Redshift Survey (da Costa 
et al. 1998).

\begin{figure*}
\epsfxsize=0.7\textwidth 
\hspace*{-0.5cm} 
\centerline{\epsffile{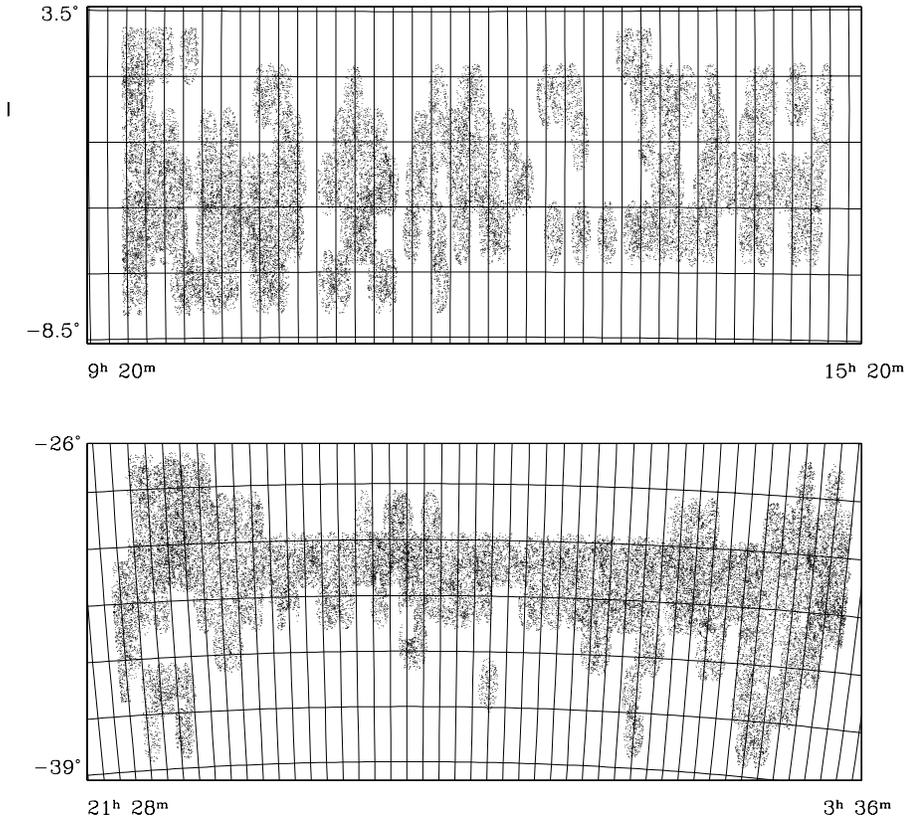}}
\caption
{Aitoff projections of the 2dFGRS. The upper projection represent
the distribution of galaxies in equatorial coordinates in  the northern 
strip and the lower the southern one.} 
\label{fig:pos}
\end{figure*}

At the present, the largest sample of galaxies with spectroscopic redshifts
($\sim 100000$) is the Anglo-Australian two degree field galaxy redshift 
survey (hereafter 2dFGRS, Colless et al. 2001).
The survey covers $2000 \ deg^2$ and has a median depth $\bar{z}=0.11$.

Several studies were done with this preliminary sample by the 2dFGRS Team.
Some of then are
the power spectrum of galaxies (Percival et al. 2001), the projected two point 
correlation function of galaxies for different volume limited samples 
(Norberg et al. 2001a), and the estimation of the $b_J$ band galaxy 
luminosity function (Norberg et al. 2001b).
Another study related with system of galaxies in the 2dFGRS has been 
carried out by De Propris et al. (2002) where they present an study of 
3-dimensional galaxy clusters based on well known bidimensional galaxy 
clusters.

The large number of galaxies and the depth of the 2dFGRS make it very
suitable for galaxy groups identification. 
Nevertheless the present release of the sample is not uniform due to 
variations in sky coverage which make difficult the analysis of the data. 

The main objective of this work is the identification of groups in the 2dFGRS
and estimation of their physical properties (number of members, velocity 
dispersion, virial radius and mass).
In order to take into account the variation in the sky coverage of the 2dFGRS
we introduce a couple of modifications to the group-finding algorithm 
developed by Huchra \& Geller (1982). The first one deals with the 
apparent magnitude limit variation across the sky and the second is
related to redshift completeness. 
These modifications are tested using mock catalogues which reproduce each 
effect separately. 
We also use these mock catalogues to explore the space of linking
parameters to maximize group identification accuracy.

The paper is structured as follows. The catalogue is described in section 2. 
In section 3 we describe the group finder algorithm used to identify galaxy 
groups in the 2dFGRS. A detailed test of the method is performed in section 4.
Finally, the 2df group catalogue is presented in section 5 followed by 
conclusions in section 6 .

\section{The galaxy catalogue}

The complete 2dFGRS will consist of approximately 250000 galaxies with 
redshifts in two declination strips plus 100 random 2-degree fields.
All targets are selected in the photometric $b_j$ band from the APM galaxy 
survey (Maddox et al. 1990a,b;1996). The southern strip (SGP) covers 
approximately 1275 square degrees 
($-37^{\circ}.5 \leq \delta \leq -22^{\circ}.5$; 
$ 21^h 40^m \leq \alpha \leq 3^h 30^m$) and the northern strip (NGP) covers 750 
square degrees ($-7^{\circ}.5 \leq \delta \leq 2^{\circ}.5$; $9^h 50^m \leq 
\alpha \leq 14^h 50^m$) while the 100 random fields are spread uniformly over 
the 7000 square degrees in the southern region.

In our analysis we use the 2dF public 100K data release of galaxies with the
 best redshift 
estimates within the two main strips (NGP and SGP) of the catalogue. 
Given the status of the project, the sky coverage of the sample
is not uniform, so a detailed completeness description is needed 
(see Figure \ref{fig:pos}).
For this purpose, we use in our analysis the 2dFGRS mask software constructed
by Peder Norberg and Shaun Cole which take into account both the magnitude 
limit and the redshift completeness (see also Figure 13 and 15 of 
Colless et al. 2001).
The redshift completeness was defined with the $2^\circ$ field used to
tile the survey region for spectroscopic observations.  
This quantity is the ratio of the number of galaxies for which redshifts have 
been obtained to the total number of objects contained in the parent catalogue
and has a mean value of $\sim 0.75$ for the whole sample.
On the other hand, the magnitude limit mask corresponds to variations of the
parent survey magnitude limit with the position on the sky. This variation 
span over a
magnitude limit range of $m=18.95-19.55$ for the NGP and a range of 
$m=19.3-19.55$ for the SGP.
This sample comprise 102426 galaxies with final $b_j$ magnitudes
corrected for galactic extinction. 

\section{The group-finding algorithm}
For group identification we use a friends-of-friends algorithm 
similar to that described by Huchra \& Geller (1982) modified in order 
to take into account redshift completeness and magnitude limit variations.

If we have a pair of galaxies with mean radial velocity $V=(V_1+V_2)/2$ and 
angular separation $\theta_{12}$, our algorithm links galaxies when the 
following conditions are satisfied:
\begin{equation}
D_{12}=2 \sin\left(\frac{\theta_{12}}{2}\right)\frac{V}{H_0} \leq D_L
\end{equation}
and 
\begin{equation}
V_{12}=|V_1-V_2| \leq V_L
\end{equation}
where $D_{12}$ is the angular separation and $V_{12}$ is the line-of-sight
velocity difference. The transverse $D_L $ and radial linking 
lengths $V_L$ are scaled as $D_0 \ R$ and $V_0 \ R$ respectively 
in order to compensate for the decline of the selection function with distance.
The scaled factor is computed as
\begin{equation}
R = \left[\frac{\int_{-\infty}^{M_{12}}\phi(M)dM}{\int_{-\infty}^{M_{lim}}\phi(M)dM}\right]^{-1/3} 
\end{equation}
where $M_{lim}$ and $M_{12}$ 
are the absolute magnitude of the brightest galaxy visible at a distance 
$V_f/H_0$ and $V/H_0$ respectively. In these equations, $D_0$ and $V_0$ are
the linking cutoffs at the fiducial velocity $V_f$, and $\phi(M)$ is the 
galaxy luminosity function of the sample.

In the special case of the 2dFGRS, we should take into account the magnitude
limit and completeness variations through the sky. 
The adopted way to solve 
this issue consists in defining an average magnitude limit for each pair
\begin{equation}
m_{lim}=(m_{lim}^1+m_{lim}^2)/2
\end{equation}  
and redefining the scale factor $R$ as
\begin{equation}
R=\left[\frac{\int_{-\infty}^{M_{12}}\phi(M)dM}{\int_{-\infty}^{M_{lim}}\phi(M)dM}\frac{(C_1+C_2)}{2}\right]^{-1/3} 
\end{equation}
where $C_1$ and $C_2$ are the corresponding completeness values for each galaxy
position on the sky.

\section{Testing the method}

\subsection{Mock Catalogues}
To examine the degree of accuracy of our algorithm we use a set of mocks 
catalogues constructed from a gravitational numerical simulation of
a flat low density cold dark matter universe. 
We perform this simulation using the Hydra Nbody code developed by 
Couchman et. al (1995)
with $128^3$ particles in a cubic comoving volume of 180 $h^{-1}$ Mpc per 
side starting at z=50.
The adopted cosmological model was a universe with $\Omega_m=0.3$, $\Omega_
{\Lambda}=0.7$, Hubble constant $h=0.7$ and a relative mass fluctuation of
$\sigma_8=0.9$.   

\begin{figure}
\epsfxsize=0.5\textwidth 
\hspace*{-0.5cm} \centerline{\epsffile{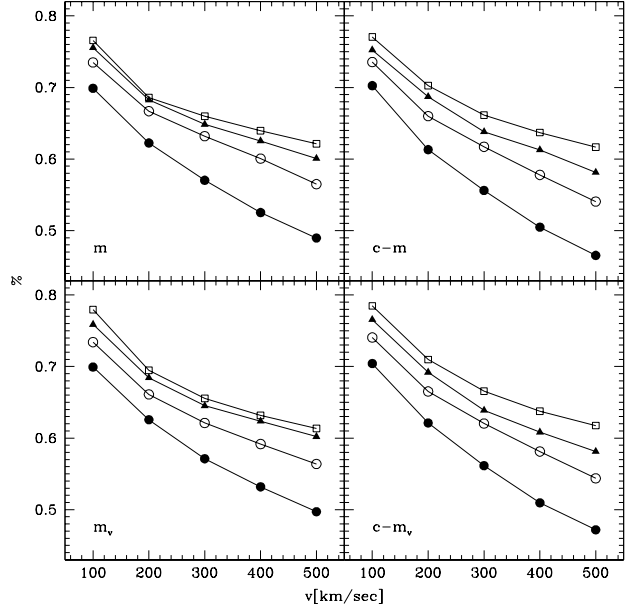}}
\caption
{The percentage of groups found in redshift space which match with some group
in real space. These percentage are shown spanning the parameter space
$V_0$ (100, 200, 300, 400 and 500 $km^{-1}$) $\delta \rho/\rho$ (20-filled
circles, 40-open circles, 80-filled triangles and 160-open squares). 
Each panel show the different kind of mocks constructed.} 
\label{fig:matchzpor}
\end{figure}

\begin{figure}
\epsfxsize=0.5\textwidth 
\hspace*{-0.5cm} \centerline{\epsffile{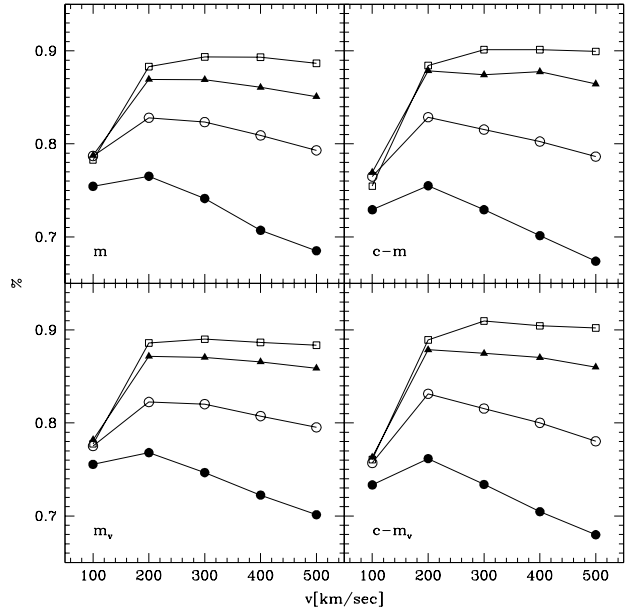}}
\caption
{The percentage of the group catalogue in real space associated with the
catalogue of matched groups in redshift space.
The symbols and the panels are the same as in Figure \ref{fig:matchzpor}.} 
\label{fig:matchzr}
\end{figure}

In order to reproduce the same radial distribution as in the 2dFGRS, we 
adopt a galaxy luminosity function fitted by a Schechter function with
$M_{b_j}^{\ast}-5log_{10}h=-19.66$, $\alpha=-1.21$, $\Phi^{\ast}=1.68\times
 10^{-2}h^3Mpc^{-3}$ and a model of the average k+e correction given by the
formula 
\begin{equation}
k(z)+e(z)=\frac{z+6z^2}{1+20z^3}
\end{equation}
(see Norberg et al. 2001b). Combining these models and using the positions
from the simulation we generate an apparent magnitude 
\begin{equation}
m=k+e+5\log_{10}(d_L/h^{-1}Mpc)+25+(M_{b_j}-5 \log_{10} h)
\end{equation}
for each particle inside the angular mask described in section 2. 
We construct four mock catalogues with different sky coverage:
\begin{itemize}
\item in the first mock we introduce a fixed faint survey 
magnitude limit ($m$); 
\item the second mock has a fixed faint survey magnitude limit and the effect
of redshift completeness as in the real survey ($c-m$);
\item the third mock has a faint survey magnitude limit which changes with the
angular position of a particle in the same way as the magnitude limit mask of
the 2dFGRS ($m_v$);
\item finally, the fourth mock have both effects, the magnitude limit mask and
the redshift completeness mask ($c-m_v$).
\end{itemize} 

\subsection{Identification Accuracy}
From the mock catalogues described in the previous section, we identify
groups in real and redshift spaces. The former identification is carried out
using the same linking length in both directions ($V_0=D_0$) whereas 
different linking lengths are used in redshift space. The fiducial velocity
$V_f$ is adopted as 1000 $km$ $s^{-1}$ in all identifications.
The linking lengths are selected spanning the space parameter ($V_0$,
$\delta \rho/\rho$) where the later is the number density
contour surrounding a group and corresponds to a fixed number density 
enhancement relative to the mean number density of
\begin{equation}
\frac{\delta \rho}{\rho}= \frac{3}{4\pi D_0^3}\left( 
\int_{-\infty}^{M_{lim}} \Phi(M) dM \right)^{-1} -1
\end{equation}
The selected values are:
\begin{itemize}
\item $V_0$ : 100, 200, 300, 400 and 500 $km$ $s^{-1}$
\item $\delta \rho/\rho$ : 20, 40, 80 and 160
\end{itemize}

To quantify how good are the groups identified in redshift space we
implement a method to match the center of mass of these groups with the
center of mass of groups in real space. 
Our method seeks for groups in real space within a projected 
and a radial distance having its origin in the center of mass of each 
group in redshift space. We choose the searching parameters in order
to maximize the matching, nevertheless the results are very stable around
the chosen values. 
The results show that almost all groups have one match inside the limits
and only a few are related with two or three groups in real space.
In the later situation we choose to associate the group in redshift space
with that group in real space with more shared particles.

\begin{figure}
\epsfxsize=0.5\textwidth 
\hspace*{-0.5cm} \centerline{\epsffile{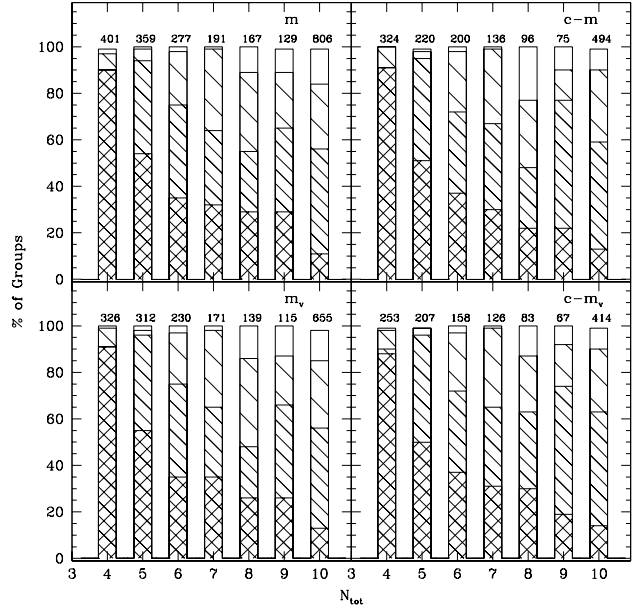}}
\caption
{
Fraction of groups of a given richness $N_{tot}$ with ${\cal F}$.
Cross hatched regions show a ratio ${\cal F}$ of unity, 
single narrow hatched regions correspond to $0.75\le{\cal F}<1$,
$0.5\le{\cal F}<0.75$ to single wide hatched regions and no hatching regions 
belong to $0.25\le{\cal F}<0.5$. 
For richness 10 we include all groups with 10 or more
members.
The labels over the bars show the number of groups for the 
corresponding richness.
} 
\label{fig:match}
\end{figure}

\begin{table}
\begin{center}
\caption{Percentage of spurious groups identified in the $c-m_v$ mock}
\begin {tabular}{ccccc}
\hline 
\hline 
$V_0$ &  \multicolumn{4}{c}{$\delta \rho/\rho$}  \\
\cline{2-5}
(km $s^{-1}$) & 20 & 40 & 80 & 160 \\
\hline 
     & 13  &  9  &  6  &  4  \\
 100 &  7  &  8  &  8  &  8  \\
     &  9  &  9  &  9  &  9  \\
\hline 
     & 16  & 10  &  6  &  5  \\
 200 & 10  & 10  & 11  & 10  \\
     & 12  & 13  & 13  & 14  \\
\hline 
     & 19  & 12  &  9  &  6  \\
 300 & 11  & 12  & 13  & 12  \\
     & 13  & 14  & 14  & 15  \\
\hline 
     & 23  & 14  & 10  &  7  \\
 400 & 13  & 13  & 14  & 14  \\
     & 13  & 15  & 15  & 15  \\
\hline 
     & 26  & 17  & 12  &  8  \\
 500 & 14  & 14  & 15  & 14  \\
     & 14  & 15  & 15  & 16  \\
\hline
\end{tabular}
\end{center}
\end{table}

The first result from this comparison is that the percentage of spurious
groups identified is remarkably higher when we include groups in redshift
space with 3 members ($\sim 40 \%$). Ramella, Geller \& Huchra (1989)
and Frederic (1995) obtain that one-third or more of the groups identified in 
redshift space with N=3 are spurious, which is roughly consistent with 
our result. 
Consequently, we adopt as the main catalogue all groups identified in 
redshift space with more than 4 members.
Figure \ref{fig:matchzpor} shows the number of groups which match
with one group in real space relative to the total number of groups in
the main catalogue. The analysis comprises all mock and identification
parameters under consideration. 
The complementary percentage in each case represents the spurious groups
in the main catalogue. We consider spurious those groups which do not have 
any match or are associated with a binary or a triplet in real space.   
The individual percentages for these spurious groups for the $c-m_v$ mock
are shown in Table 1 where the first line represents no matched groups, the 
second and third lines correspond to binaries and triplets respectively.

\begin{figure*}
\epsfxsize=.79\textwidth 
\vspace{-2cm}
\hspace*{-0.5cm} 
\centerline{\epsffile{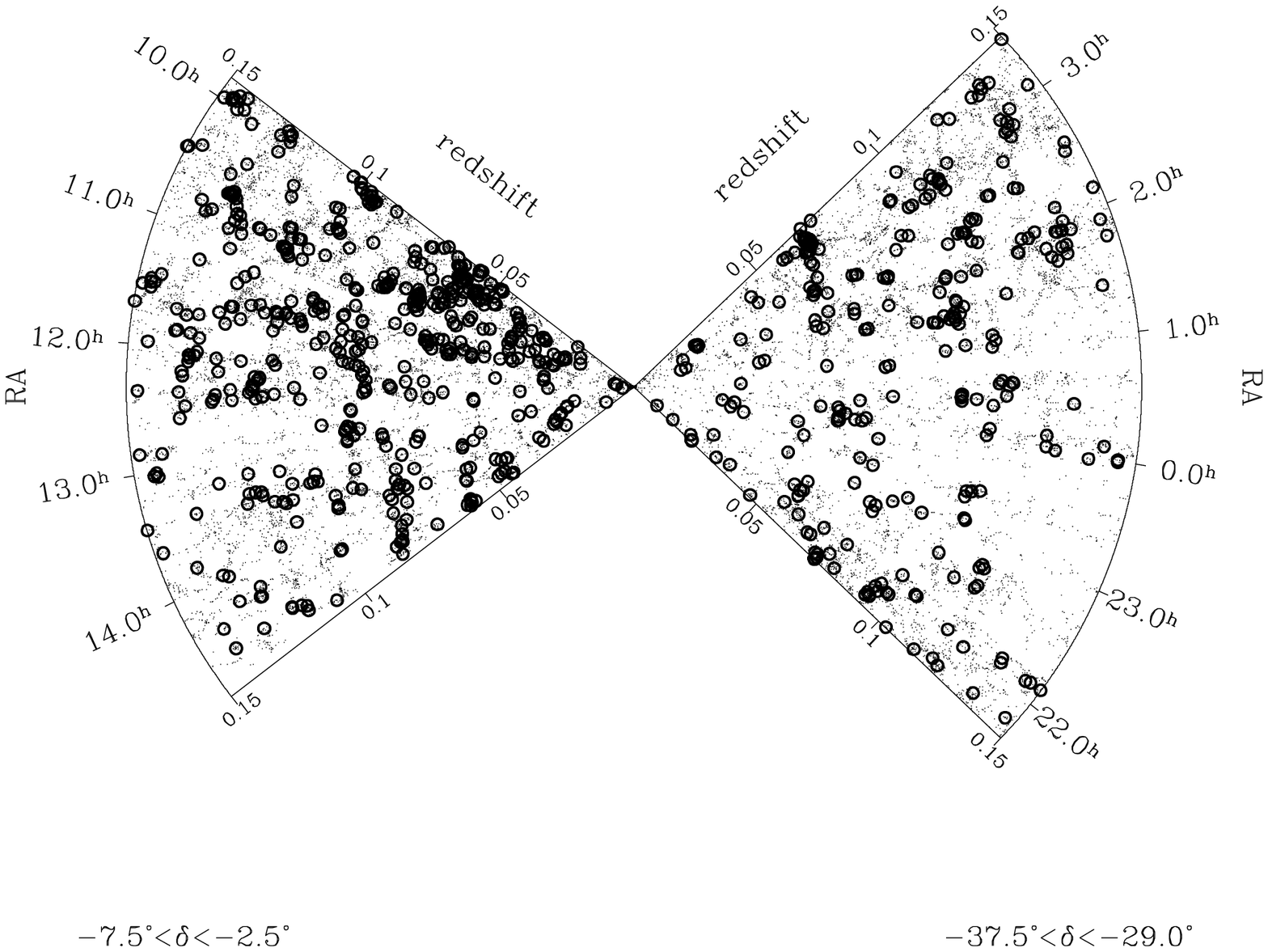}}\\
\vspace{-3cm}
\hspace*{-0.5cm} 
\centerline{\epsffile{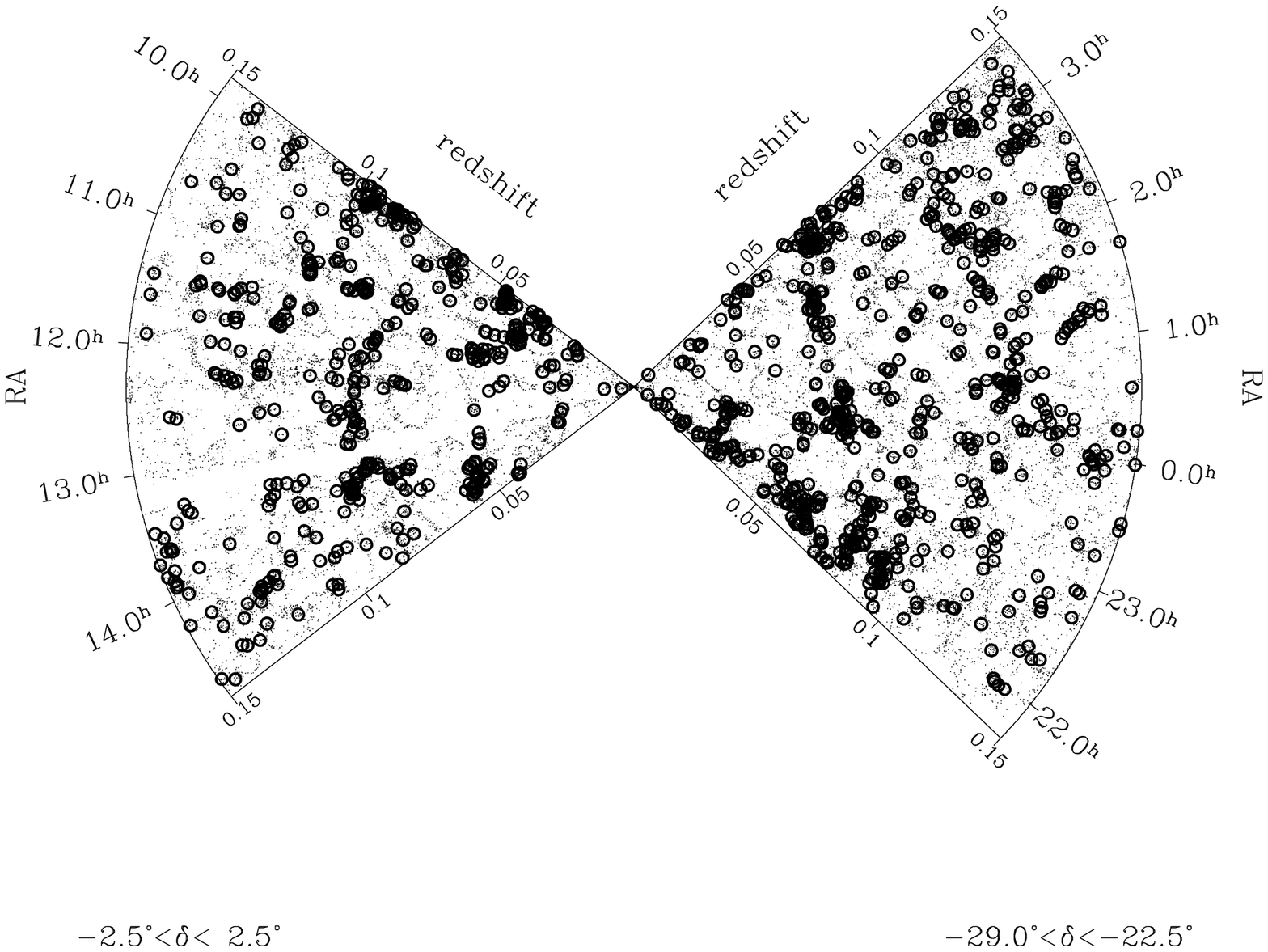}}
\caption
{The redshift distribution for the galaxy groups identified in the 2dFGRS.
The pieplot show at the left the NGP of the 2dFGRS and the SGP at right.
The dots are the galaxies and the open circles the galaxy groups.
To make the plot clearer we limit the plot to $z<0.15$ } 
\label{fig:pie2df}
\end{figure*}

We also analise the percentage of the group catalogue in real 
space associated with matched groups in the main catalogue. 
That percentage is plotted in Figure \ref{fig:matchzr} for all the options 
as in Figure \ref{fig:matchzpor}. 
Despite the percentage of matched groups in the main catalogue 
seems to be the highest for a velocity of 100 $km$ $s^{-1}$ and
density contrast of 160, it can be seen from Figure \ref{fig:matchzr}
that this choice represents one of the lowest percentages of the group
catalogue in real space.
Combining the two figures we can conclude that the best options are a 
velocity of 200 $km$ $s^{-1}$ and density contrast of 80 or 160. 
We choose a density contrast of 80 because the total number
of groups identified is greater and the matching accuracy is roughly similar.

\begin{figure}
\epsfxsize=0.5\textwidth 
\hspace*{-0.5cm} 
\centerline{\epsffile{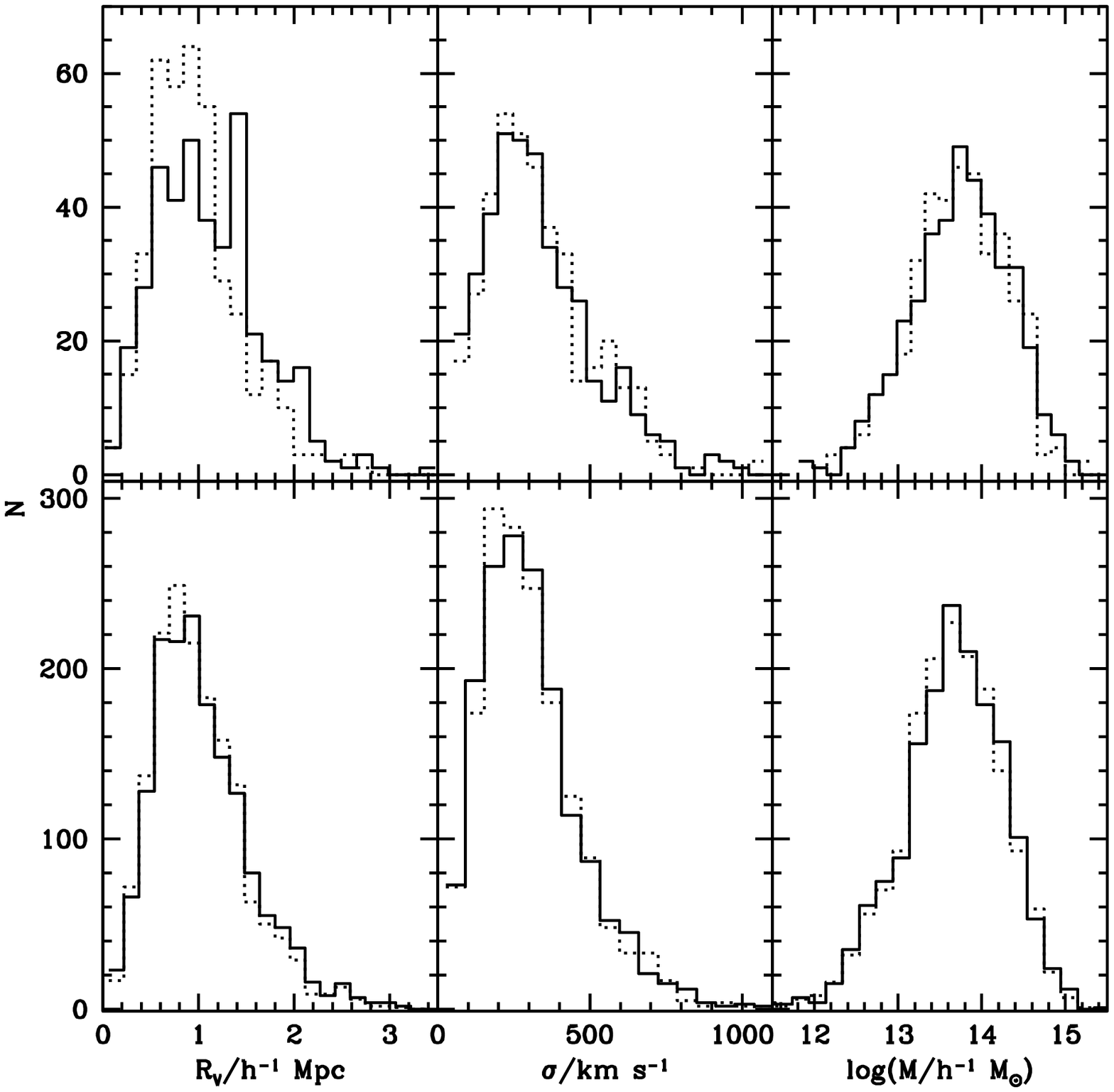}}
\caption
{ Upper and lower panels show the distribution of the main physical properties
of groups at low and high completeness regions respectively. Solid and doted
lines correspond to groups with and without border effects.
} 
\label{fig:comp}
\end{figure}

The resulting value of $\delta \rho/\rho =80$ is in agreement with that  
obtained by Ramella et al. (1997) for the CfA2 Redshift Survey, whereas our 
optimal velocity linking parameter $V_0=200 \ km \ s^{-1}$ differs from his 
choice ($V_0=350 \ km \ s^{-1}$). Should be taken into account that equal 
$\delta \rho/\rho$ correspond to different values of $D_0$ depending
of the apparent magnitude limit of the survey. Particularly, the
the apparent magnitude limits for the 2dFGRS and the CfA2 Redshift Survey
are $\sim 19.4$ and $15.5$ respectively. Consequently, the higher number
density of the 2dFGRS implies smaller linking parameters than those obtained
for the CfA2 Redshift Survey.

From the comparison between the four kind of mock catalogues we conclude
that implemented modifications keep the behavior of the original algorithm,
showing the ability of these modifications to deal with an inhomogeneous 
galaxy samples like the 2dFGRS.

To examine the degree of agreement between members of groups in real and 
redshift space we perform an analysis similar to Frederic (1995) for matched
groups in the main catalogue.
If we define the total number of members of a given group in redshift
space as $N_{tot}$, we search how many members of this group are in the
associated group in real space $N_m$ and compute the ratio 
${\cal F}=N_m/N_{tot}$.  Figure \ref{fig:match} shows the fraction of 
groups of a given richness $N_{tot}$ with ${\cal F}$ for several values:
1 (cross hatched regions), $0.75\le{\cal F}<1$ (single narrow hatched regions),
$0.5\le{\cal F}<0.75$ (single wide hatched regions) and $0.25\le{\cal F}<0.5$
(no hatching regions). For richness 10 we include all groups with 10 or more
members. As it can be seen, more than $\sim$65\% of the groups have a 
ratio ${\cal F}$ greater than 0.75 for all kind of mocks. It also should be 
noticed that the implemented modifications keep the performance of the original 
Huchra \& Geller algorithm. 

We still need to test that our modifications to the Huchra \& Geller algorithm
improve the detection of galaxy groups. In order to prove the modification
which take into account the redshift completeness we compare the results of
the identification using both, modified and original Huchra \& Geller 
algorithms on the $c-m$ mock catalogue. As result, we find that the 
identification using our modification produce $263$ ($\sim 11\% $ 
of the total sample) more groups than the
corresponding to the plain algorithm version. 
On the other hand, the wide apparent magnitude limit range implies a variation
in the local number density, which produce a similar effects as the redshift
completeness. This effect was tested applying a similar procedure as the former
but using the $m_v$ mock catalogue. 
The results show a difference of $89$ groups using our modified algorithm
against the original Huchra \& Geller one.    
The effect due to the apparent magnitude limit is smaller than the
observed for the redshift completeness because the magnitude limit distribution
have a mean value of $\sim 19.4$ tending to the upper limit of the whole
range. 
Finally, we test the joint effects applying the same procedure to the
$c-m_v$ mock catalogue, showing that our modification identify $280$ more 
groups ($\sim 14\%$ of the total sample). 
As we have shown in the previous tests, the modified algorithm keep the 
performance
of the original Huchra \& Geller method, then the $ \sim 70\% $ of these 
differences correspond to real groups for the chosen linking parameters.

\section{Galaxy group identification in the 2dFGRS}

To identify groups in the 2dF survey (2dFGGC), we adopt the values 
$\delta \rho/\rho=80$ and $V_0=200$ $km$ $s^{-1}$ 
which maximize the group accuracy as shown in section 4. 

The resulting groups catalogue contains systems with at least 4 members, 
mean radial velocities in the range $900 km s^{-1}\leq V \leq 75000 km s^{-1}$
and a total number of 2209 groups. 
It represents the largest sample to the present and provides a suitable data 
set to analise the clustering properties of galaxy systems of low richness. 
As discussed in the previous section, the limit adopted in the number of 
members in galaxy groups is necessary in order to avoid pseudo-groups.

Figure \ref{fig:pie2df} shows the cone diagrams for galaxies (points) and 
groups (open circles) in the 2dF survey.

\begin{figure}
\epsfxsize=0.5\textwidth 
\hspace*{-0.5cm} \centerline{\epsffile{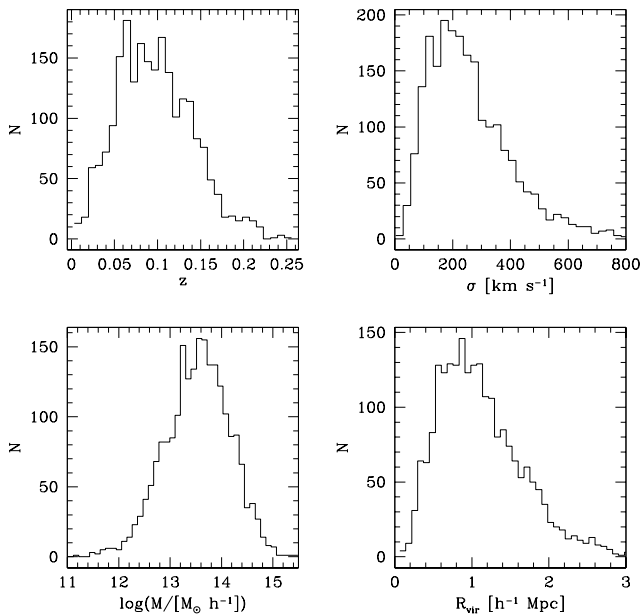}}
\caption
{The upper-left panel shows the histogram describing the radial distribution for
the groups identified in the 2dFGRS. The upper-right panel shows the 
distribution of groups velocity dispersions.
The lower-left panel plots the virial
mass distribution whereas the lower-right panel shows the distribution of 
groups virial radii.}
\label{fig:zr}
\end{figure}

Estimation of the virial mass for galaxy groups is carried out using the
following equation 
\begin{equation}
M_{vir}=\frac{\sigma^2 R_V}{G}
\end{equation}
where $R_V$ is the virial radius of the system and $\sigma$ is the 
velocity dispersion of member galaxies (Limber \& Mathews 1960).
The virial radius is estimated as
\begin{eqnarray}
R_V & = & \frac{\pi}{2}R_{PV} \\
R_{PV}& = & \frac{N_g(N_g-1)}{\sum_{i>j}R_{ij}^{-1}}
\end{eqnarray}
where $R_{PV}$ is the projected virial radius, $N_g$ is the number of galaxies
members and $R_{ij}$ the galaxy projected distances. 
The velocity dispersion $\sigma$ is estimated using their observational 
counterpart, the line-of-sight velocity dispersion $\sigma_{v}$,
$\sigma=\sqrt{3}\sigma_{v}$. 
In particular we use the methods described by
Beers, Flynn \& Gebhardt (1990) to obtain a robust estimation of the 
radial velocity dispersion. We apply the biweight estimator for groups
with richness $N_{tot}\ge 15$ and the gapper estimator for poorer groups
(Girardi et al. 1993, Girardi \& Giuricin 2000). 
Despite these methods improve the velocity dispersion estimation in terms of 
efficiency and stability when dealing with small groups, it should be taken 
into account that small errors in the estimate of the velocity dispersion 
results in correspondingly larger errors in the derived mass.

\begin{table}
\tiny
\begin{center}
\caption{Mean parameters of the 2dFGGC}
\begin {tabular}{ccccc}
\hline 
$N$ & $\bar{V_r}(km/s)$ & $\bar{\sigma}(km/s)$ & $\bar{M}(h^{-1} M_{\odot})$ & $\bar{R}_V(h^{-1}Mpc)$\\
\hline 
$2209$ & $31500$ & $261$ & $8.5\times 10^{13}$ & $1.12$\\

\hline
\end{tabular}
\end{center}
\end{table}

The group properties estimation could be biased by the geometrical and 
sampling effects in the 2dF catalog.
To deal with border effects we create a simple magnitude limited mock catalogue
where limits in declination and right ascension exceed the corresponding to 
the 2dF catalog. 
Consequently we identify groups in this mock catalogue keeping those 
inside the 2dF border and we confront them with groups identified
in a mock catalogue which emulate the 2dF geometry. As result, we observe no 
differences between the distribution of physical properties 
($R_V$,$\sigma$ and $M$) of these samples, showing that the complex geometry
of the 2dF border do not affect group properties.

Another source of bias could be the variation of redshift completeness 
across the sky producing a wrong estimation of the physical properties
in low completeness regions. In order to quantify this possible bias, we
compare the physical properties of both, $c-m$ and $m$ mock catalogues
described in section 4, splitting these samples in high and low redshift
completeness. In Figure \ref{fig:comp} we show the histogram of $R_V$,
$\sigma$ and $M$ for low ($<0.6$,upper panels) and high ($\ge0.6$,lower 
panels) completeness. Dotted lines correspond to $m$ mock catalogue
whereas the solid lines correspond to $c-m$ one. As it can be seen,
only the virial radius in the low redshift completeness regions
show some differences, nevertheless they do not produce any significant effect
in the mass estimation.

Finally the distribution of the estimated physical properties are shown in 
Figure \ref{fig:zr}.
In the upper-left panel we plot the redshift 
distribution with a mean redshift of $z \sim 0.1 $,  the lower-left panel 
shows the virial mass distribution, the upper-right panel correspond to
the velocity dispersion distribution, and the virial radii 
distribution is shown in the remainder plot.
The mean values of the 2dFGGC are quoted in Table 1.

\section{CONCLUSIONS}
In this paper we present a new catalogue of galaxy groups derived from 
the 2dFGRS,
the largest sample of galaxies with spectroscopic redshifts at the 
present.  The construction of this catalogue of groups (2dFGGC) was carried out 
introducing some modifications in the 
friend of friends algorithm developed by Huchra \& Geller (1982)
in order to take into account the different sky coverage variations
in the sample.
We have tested these modifications taking advantage of mock catalogues 
which allow us to explore the space parameter looking for the best values to
maximize the group identification accuracy in redshift space. 
With these parameters ($\delta \rho/\rho=80$ and $V_0=200$ $km$ $sec^{-1}$) 
we obtain that $\sim 90 \%$ of groups in real space match some group 
with at least four members in redshift space. Furthermore, the $\sim 30 \%$
of groups in redshift space are spurious (no matched groups $\sim 6 \%$
and groups associated with a binary or triplet $\sim 24 \%$). 
We also find that the introduced modifications are able to detect $\sim 14\%$ 
more groups than the obtained by the original Huchra \& Geller algorithm. 
Additionally, we explore the border effect and a possible bias due to
redshift completeness on the physical properties, finding
that none of them produce significant effects on the final estimations.

Using the optimal identification parameters, the final group sample for the 2dF
survey comprise 2209 groups of galaxies with mean  
redshift of $\bar{z} \sim 0.1$ and a mean velocity dispersion of
$\bar{\sigma}$=261 km/s.
From the estimation of the virial masses for the galaxy groups we obtain  
a mean mass $\bar{M}=$ $8.5 \times 10^{13}$ $h^{-1}$ $M_{\odot}$  with
a mean virial radius $\bar{R_V}=1.12$ $h^{-1}\ Mpc$.

The new galaxy group catalogue
is one of the largest groups samples until the present. 
For this reasons the new sample is more suitable for statistical studies than 
has been previously available. 

\vspace{0.5cm}


\section*{Acknowledgments}
We want warmly thank to Carlton Baugh for reading the manuscript and 
helpful suggestions.
We also thank to Peder Norberg and Shaun Cole for kindly providing the 
software describing the mask of the 2dFGRS and to the 2dFGRS Team
for having made available the actual data sets of the sample.
This work has been partially supported by Consejo de Investigaciones 
Cient\'{\i}ficas y T\'ecnicas de la Rep\'ublica Argentina (CONICET), the
Secretar\'{\i}a de Ciencia y T\'ecnica de la Universidad Nacional de C\'ordoba
(SeCyT), the Consejo de Investigaciones Cient\'{\i}ficas y T\'ecnol\'ogicas de 
la  Provincia de C\'ordoba (CONICOR) and Fundaci\'on Antorchas, Argentina.

\end{document}